\documentclass[aps,pre,twocolumn,groupedaddress]{revtex4}

\usepackage{graphicx}
\usepackage{amsmath}
\usepackage{amssymb}
\usepackage{amsthm}
\usepackage{array}
\usepackage{xy}
\usepackage{subfigure}

\begin{document}


\title{Ratio of effective temperature to pressure controls the mobility of sheared hard spheres}


\author{Thomas K. Haxton}
\affiliation{The Molecular Foundry, Lawrence Berkeley National Laboratory, Berkeley, CA, 94720}
\affiliation{Department of Physics and Astronomy, University of Pennsylvania, Philadelphia, PA, 19104}


\date{\today}

\begin{abstract}
Using molecular dynamics simulation, we calculate fluctuations and response for steadily sheared hard spheres over a wide range of packing fractions $\phi$ and shear strain rates $\dot\gamma$, using two different methods to dissipate energy.  To a good approximation, shear stress and density fluctuations are related to their associated response functions by a single effective temperature $T_{\rm eff}$ that is equal to or larger than the kinetic temperature $T_{\rm kin}$.  We find a crossover in the relationship between the relaxation time $\tau$ and the the nondimensionalized effective temperature $T_{\rm eff}/p\sigma^3$, where $p$ is the pressure and $\sigma$ is the sphere diameter.  In the \textit{solid response} regime, the behavior at fixed packing fraction satisfies $\tau\dot\gamma\propto \exp(-cp\sigma^3/T_{\rm eff})$, where $c$ depends weakly on $\phi$, suggesting that the average local yield strain is controlled by the effective temperature in a way that is consistent with shear transformation zone theory.  In the \textit{fluid response} regime, the relaxation time depends on $T_{\rm eff}/p\sigma^3$ as it depends on $T_{\rm kin}/p\sigma^3$ in equilibrium.  This regime includes both near-equilibrium conditions where $T_{\rm eff} \simeq T_{\rm kin}$ and far-from-equilibrium conditions where $T_{\rm eff} \ne T_{\rm kin}$.  We discuss the implications of our results for systems with soft repulsive interactions.
\end{abstract}


\maketitle

\section{Introduction}

Many applications require understanding how disordered materials flow under an external load such as a shear stress~\cite{Coussot2007, Brader2010}.  Deriving such an understanding from thermodynamic principles requires identifying relationships between thermodynamic parameters and transport.  Since external loads drive systems out of equilibrium, their thermodynamics cannot be described solely in terms of equilibrium parameters like temperature and pressure.  However, simulations~\cite{Barrat2000, Berthier2002, Berthier2002b, Ono2002, Makse2002, OHern2004, Ilg2007} and experiments~\cite{Corwin2005, Song2005, Wang2008} show that sheared spherical particles possess an \textit{effective} temperature that relates low-frequency fluctuations of various observable quantities to their associated response functions.  This effective temperature is likely to feature prominently in any thermodynamics-based description of flow in glassy materials~\cite{Falk2011}.  First, the relationship between transport and effective temperature must be established.

In Ref.~\cite{Haxton2007} we begun to establish a relationship between mobility and effective temperature $T_{\rm eff}$ by studying the behavior of a mixture of soft, repulsive disks in two dimensions at a packing fraction above random close packing over a range of shear strain rates $\dot\gamma$ and kinetic temperatures $T_{\rm kin}$.  We found that the system only flows under shear if $T_{\rm eff}$ exceeds a threshold value similar to the dynamic glass transition temperature.  Under strong shearing, we found that the average shear stress is proportional to a Boltzmann factor containing the effective temperature.  These results imply that the effective temperature plays a key role in facilitating the mobility of sheared systems.

Here, in order to further elucidate this role, we determine the relationship between mobility and $T_{\rm eff}$ over a wide range of parameters for a hard sphere model.  The advantages of the hard sphere model are twofold.  First, due to the hard-core nature of the interactions, $T_{\rm eff}$ can only control the behavior of the model via a dimensionless parameter.  Using the dimensionless parameter $T_{\rm eff}/p\sigma^3$, where $p$ is the pressure and $\sigma$ is the sphere diameter, facilitates a direct comparison with the equilibrium system, where $T/p\sigma^3$ is the relevant control parameter.  Second, the relationship between mobility and $T_{\rm eff}/p\sigma^3$ should capture the leading order behavior of soft repulsive systems, provided that parameters continue to be expressed in dimensionless form~\cite{Xu2009b, Haxton2011, Schmiedeberg2011}.  

We find a crossover in the relationship between $T_{\rm eff}$ and the relaxation time $\tau$ that characterizes the mobility.  In the \textit{solid response} regime, $\tau$ depends on $T_{\rm eff}$ at fixed packing fraction according to $\tau\dot\gamma\propto \exp(-cp\sigma^3/T_{\rm eff})$, where $c$ depends weakly on $\phi$, suggesting that the average local yield strain is proportional to a Boltzmann factor containing the effective temperature.  This is consistent with the effective temperature's role in shear transformation zone theory, where $T_{\rm eff}$ controls the density of zones that are susceptible to shear deformation~\cite{Falk2011}.  In the \textit{fluid response} regime, the relaxation time depends on $T_{\rm eff}/p\sigma^3$ as it depends on $T_{\rm kin}/p\sigma^3$ in equilibrium.  This regime includes both near-equilibrium conditions where $T_{\rm eff} \simeq T_{\rm kin}$ and far-from-equilibrium conditions where $T_{\rm eff} \ne T_{\rm kin}$.  This suggests that the mechanisms that control transport for the equilibrium fluid persist under shear even where the slow degrees of freedom responsible for transport fall out of equilibrium with fast degrees of freedom.

\section{Model and methods}

Our model is a mixture of 4096 hard spheres of mass $m$, half each of diameter $\sigma$ and $1.4\sigma$ to avoid crystallization.  We conduct event-driven molecular dynamics simulations~\cite{Marin1993, Isobe1999} at fixed packing fraction $\phi$ and uniform shear strain rate $\dot\gamma$ under Lees-Edwards boundary conditions.  Maintaining such a system in steady state requires dissipating energy.  We perform two separate sets of simulations using two different methods to dissipate energy.  For \textit{inelastic} simulations, we dissipate energy at each collision by imposing a coefficient of restitution $C<1$.  For \textit{thermostatted} simulations, we let collisions be purely elastic ($C=1$) but dissipate energy into a thermal reservoir at temperature $T_{\rm kin}$ by periodically rescaling the velocities $\vec{v}_i$ to keep the kinetic temperature $m\langle |\vec{v}_i|^2\rangle_i/3$ within $1\%$ of $T_{\rm kin}$.  Although there are three control parameters for each set of simulations ($\phi$, $\dot\gamma$, and $C$ or $T_{\rm kin}$), one of them can be absorbed into the unit of time; for instance, time can be measured in units of $\dot\gamma^{-1}$ or $\sigma\sqrt{m/T_{\rm kin}}$.  There are therefore only two independent, dimensionless control parameters, the packing fraction $\phi$ and a parameter that controls the strength of shearing.  While $\dot\gamma\sigma\sqrt{m/T_{\rm kin}}$ and $C$ are both valid dimensionless control parameters, we choose to represent the strength of shearing in both sets of simulations by the ratio $\Sigma/p$ of shear stress to pressure.  For the thermostatted and inelastic simulations, $\Sigma/p$ can be increased at fixed $\phi$ by increasing the shear strain rate $\dot\gamma\sigma\sqrt{m/T_{\rm kin}}$ or decreasing $C$, respectively; the equilibrium limit $\Sigma/p\rightarrow 0$ corresponds to $\dot\gamma\sigma\sqrt{m/T_{\rm kin}}\rightarrow 0$ and $C\rightarrow 1$.

We characterize the mobility by the relaxation time $\tau$ defined by $\Delta r_z(\tau)=\sigma/\sqrt{3}$, where $\Delta r_z(t) \equiv \sqrt{\langle (r_z(t)-r_z(0))^2\rangle}$ is the root-mean-squared displacement in the direction perpendicular to the shear plane.

As we will discuss in the following section, we define the effective temperature $T_{\rm eff}$ as the value that replaces the temperature $T$ to satisfy two independent fluctuation-dissipation relations.  The first fluctuation-dissipation relation is the compressibility equation relating the compressibility to the amplitude of density fluctuations,
\begin{equation}
T\chi_T=\dfrac{S(0)}{\rho}=\dfrac{1}{\rho}\lim_{k\rightarrow 0}S(\vec{k}).
\label{compressibility}
\end{equation}
We define the isothermal compressibility out of equilibrium as the compressibility at fixed $T_{\rm kin}$ and $\Sigma/p$,
\begin{equation}
\chi_{T}=\dfrac{1}{\phi}\left.\dfrac{\partial\phi}{\partial p}\right|_{T_{\rm kin}, \frac{\Sigma}{p}},
\end{equation}
which we calculate by composing partial derivatives and taking finite differences.  For a two-component fluid like ours the structure factor $S(\vec{k})$ in Eq.~\ref{compressibility} is the combination of partial structure factors~\cite{Ashcroft1967}
\begin{equation}
S(\vec{k})=\frac{S_{11}(\vec{k})S_{22}(\vec{k})-(S_{12}(\vec{k}))^2}{x_1S_{22}(\vec{k})+x_2S_{11}(\vec{k})-2\sqrt{x_1x_2}S_{12}(\vec{k})},
\end{equation}
where $x_\alpha=1/2$ is the fraction of spheres of component $\alpha$.  We take the limit in Eq.~\ref{compressibility} by calculating $S(k\hat{z})$ for $2\pi kL=1, 2, ...$, where $\hat{z}$ is the direction perpendicular to shearing and $L$ is the periodic box length, and fitting the lowest-order expansion $\tilde{S}(k\hat{z})=\tilde{S}(0)+ck^2$ over the domain $k<0.3/\sigma$.

The second fluctuation-dissipation relation is the Einstein-Helfand relation~\cite{Helfand1960, Alder1970, Garcia-Rojo2006} relating the shear viscosity $\eta\equiv\Sigma/\dot\gamma$ to the amplitude of shear stress fluctuations,
\begin{equation}
\eta=\dfrac{1}{T}\lim_{t\rightarrow\infty}\dfrac{dH(t)}{dt},
\label{helfand}
\end{equation}
where 
\begin{equation}
H(t)=\dfrac{1}{2V}\left\langle|G_\eta(t)-G_\eta(0)-\Sigma Vt|^2\right\rangle,
\end{equation}
$G_\eta=\sum_i m\dot x_iy_i$, and $V$ is the volume.  The Einstein-Helfand relation is the analog of the Green-Kubo relation
\begin{equation}
\eta=\dfrac{V}{T}\int_0^\infty dt \langle\delta\Sigma(t)\delta\Sigma(0)\rangle
\end{equation}
for systems with discontinuous potentials.  We calculate the limit in Eq.~\ref{helfand} by fitting the slope of $H(t)$ vs $t$ over the range $2\tau<t<20\tau$, where $\tau$ is the relaxation time, so that we sample low-frequency fluctuations.  We find that we must average over very long simulations to converge to the equilibrium expression, Eq.~\ref{helfand}, for simulations with purely elastic collisions and no shearing.  Using this convergence as a guide, we only compute the right side of Eq.~\ref{helfand} for simulations of duration at least $500\tau$.

\section{Sheared hard spheres possess an effective temperature}

\begin{figure}
\begin{center}
\includegraphics[width=3in]{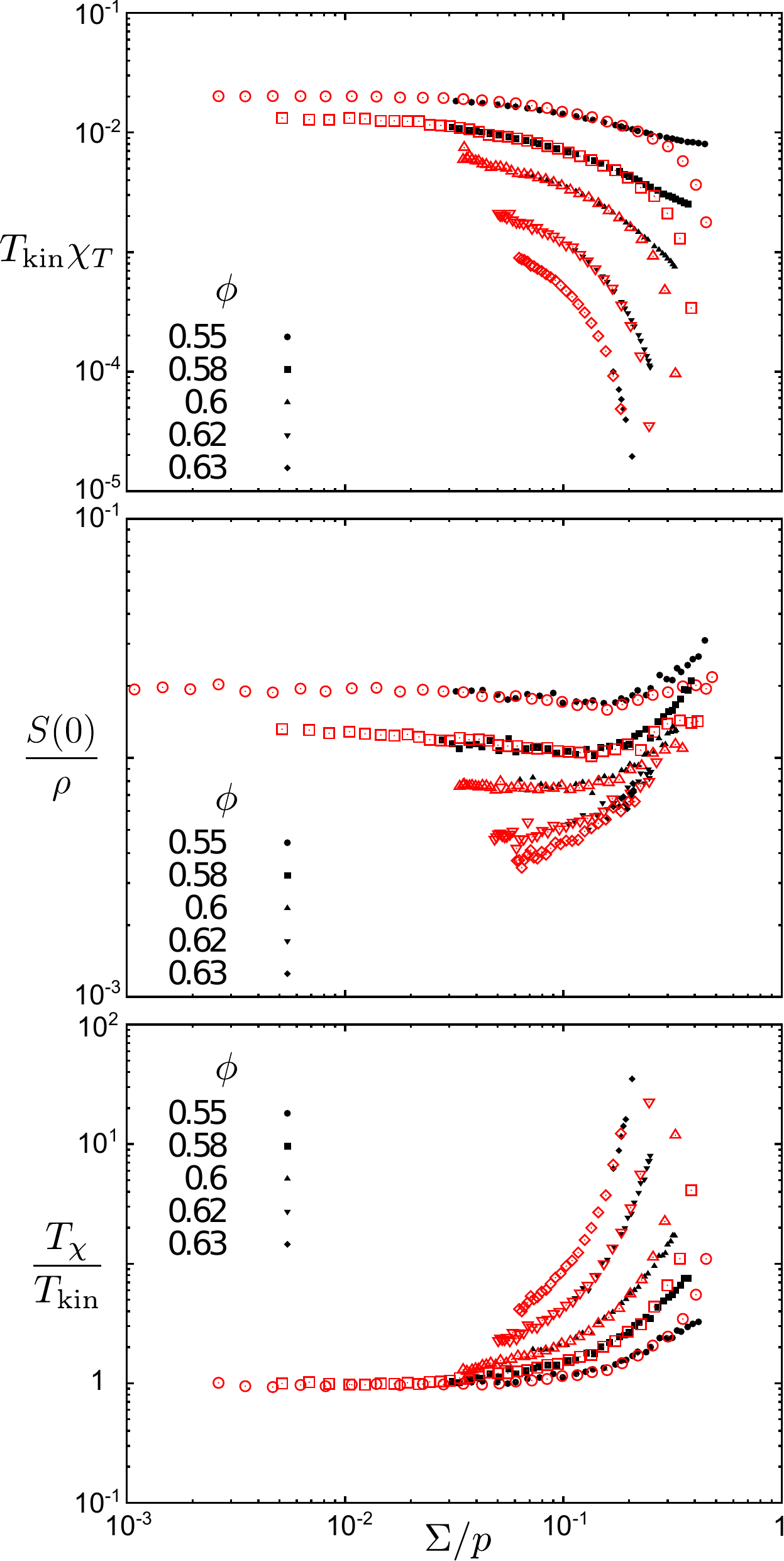}
\end{center}
\caption{(a) Measured compressibility $T_{\rm kin}\chi_T$ vs shear stress $\Sigma/p$.  (b) Compressibility expected from applying the compressibility equation (Eq.~\ref{compressibility}), $S(0)/\rho$, vs $\Sigma/p$.  (c) Ratio of compressibility effective temperature to kinetic temperature, $T_\chi/T_{\rm kin}$, vs $\Sigma/p$, where $T_\chi$ is defined via Eq.~\ref{tchi} as the ratio of the compressibilities in panels (b) and (a).  Data are presented for a selection of four packing fractions.  Red, open symbols are for thermostatted simulations, while black, filled symbols are for inelastic simulations.
}
\label{compress}
\end{figure}

Before analyzing the relationship between $\tau$ and $T_{\rm eff}$, we first show that the effective temperature is a valid concept for sheared hard spheres by demonstrating that the two independent fluctuation-dissipation relations are satisfied by a single value of $T_{\rm eff}$.  

We first consider the compressibility equation, Eq.~\ref{compressibility}, relating the isothermal compressibility to the amplitude of density fluctuations, quantified by the long-wavelength limit of the structure factor.  In Fig.~\ref{compress} (a) and (b) we plot the left and right sides of  Eq.~\ref{compressibility} for both types of energy dissipation and five selected packing fractions as a function of $\Sigma/p$.  In plotting the left side of Eq.~\ref{compressibility} in Fig.~\ref{compress} (a), we replace $T$ by the kinetic temperature $T_{\rm kin}$.  Two important features are apparent in Fig.~\ref{compress} (a) and (b).  First, except at very high $\Sigma/p$, the data for the inelastic and thermostatted simulations collapse.  With one exception that we will discuss, we find such a collapse for all observable quantities that we measure, indicating that the behavior of the model is insensitive to the way that energy is dissipated.  The breakdown of this collapse at very high $\Sigma/p$ is due to an unphysical layering transition that occurs for sheared spheres coupled to a profile-biased thermostat~\cite{Haxton2011, Evans1986, Delhommelle2003}.  At these values of $\Sigma/p$, the data for the inelastic simulations should be considered the physically realistic branch.

The second important feature of Fig.~\ref{compress} (a) and (b) is that the data in panel (a) and (b) are not identical, indicating that the compressibility equation does \textit{not} hold under strong shear.  In particular, while the measured compressibility at fixed $\phi$ decreases with increasing $\Sigma/p$, the compressibility expected from applying the compressibility equation to the measured density fluctuations increases with increasing $\Sigma/p$.  Nevertheless, following the procedure practiced in previous studies of effective temperature~\cite{Barrat2000, Berthier2002, Berthier2002b, Ono2002, Makse2002, OHern2004, Ilg2007, Haxton2007, Corwin2005, Song2005, Wang2008}, we can define a temperature-like parameter $T_\chi$ as the quantity that replaces $T$ to satisfy Eq.~\ref{compressibility}:
\begin{equation}
T_\chi\equiv \dfrac{S(0)}{\rho \chi_T}.
\label{tchi}
\end{equation}
Fig.~\ref{compress} (c) shows the ratio of this compressibility temperature to the kinetic temperature for the five packing fractions as a function of $\Sigma/p$.  For packing fractions $\phi=0.55$ and 0.58 below the colloidal glass transition~\cite{Prasad2007}, we can easily conduct simulations in the near-equilibrium regime, where $\Sigma/p$ is small and $T_\chi \simeq T_{\rm kin}$.  As $\Sigma/p$ increases, $T_\chi$ becomes larger than $T_{\rm kin}$, indicating that the degrees of freedom associated with compression are no longer in equilibrium with the degrees of freedom associated with velocity fluctuations.  For packing fractions $\phi=0.6$, 0.62, and 0.63 above the colloidal glass transition, all of our simulations are strongly nonequilibrium, with $T_\chi \gg T_{\rm kin}$.  For each packing fraction, the left edge of the data represents the lowest strain rate $\dot\gamma\sigma\sqrt{m/T_{\rm kin}}$ (or highest coefficient of restitution $C$) that we can access on the time scale of our simulations.  Obtaining data for lower strain rates or higher restitutions of coefficient would require an intractably large separation between the time scales for collisions and shearing.

\begin{figure}
\begin{center}
\includegraphics[width=3in]{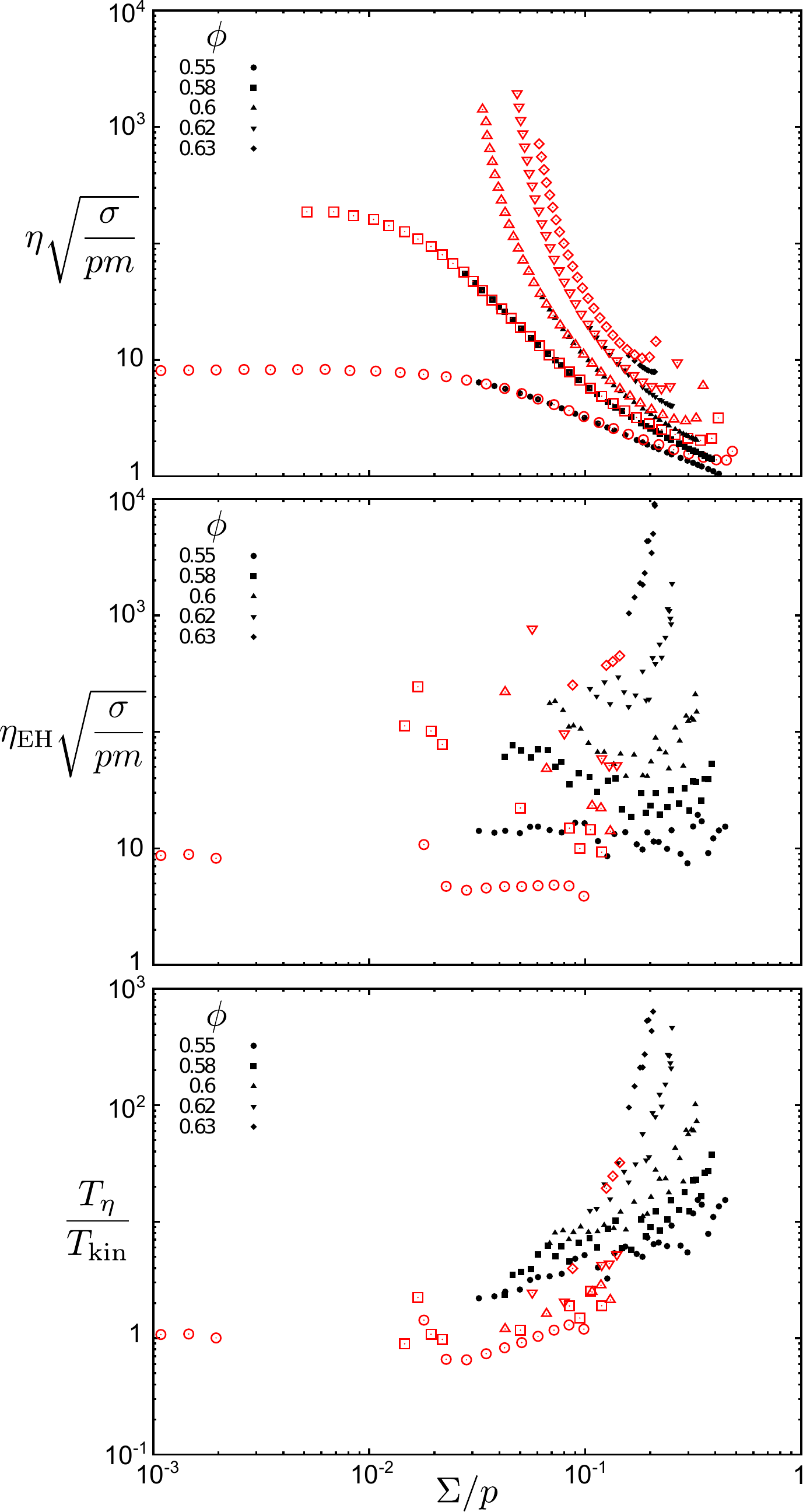}
\end{center}
\caption{(a) Measured shear viscosity $\eta\sqrt{\sigma/pm}$ vs shear stress $\Sigma/p$.  (b) Einstein-Helfand shear viscosity $\eta_{\rm EH}\sqrt{\sigma/pm}$ vs $\Sigma/p$.  (c) Ratio of viscosity effective temperature to kinetic temperature, $T_\eta/T_{\rm kin}$, vs $\Sigma/p$, where $T_\eta$ is defined by the ratio of the viscosities in panels (a) and (b).  Data are presented for a selection of four packing fractions.  Red, open symbols are for thermostatted simulations, while black, filled symbols are for inelastic simulations.
}
\label{visc}
\end{figure}

In order for $T_\chi$ to be an effective temperature, it should satisfy more than one independent fluctuation-dissipation relation.  The second fluctuation-dissipation relation that we consider is the Einstein-Helfand expression, Eq.~\ref{helfand}, relating the shear viscosity and the amplitude of shear stress fluctuations.  We define the Einstein-Helfand viscosity as the right side of Eq.~\ref{helfand}, replacing $T$ with $T_{\rm kin}$:
\begin{equation}
\eta_{\rm EH}\equiv\dfrac{1}{T_{\rm kin}}\lim_{t\rightarrow\infty}\dfrac{dH(t)}{dt}.
\label{etahelfand}
\end{equation}
In Fig.~\ref{visc} (a) and (b) we plot $\eta$ and $\eta_{\rm EH}$, respectively, as functions of $\Sigma/p$, again for both types of energy dissipation and five selected packing fractions.  We nondimensionalize the viscosities in Fig.~\ref{visc} (a) and (b) by dividing them by a characteristic viscosity that is a product of the energy density $p$ and the time scale $\sqrt{m/p\sigma}$; choosing such a characteristic energy density and time scale constructed from the pressure allows direct comparisons with systems with soft repulsions~\cite{Xu2009b, Haxton2011, Schmiedeberg2011}, but our evaluation of the effective temperature is independent of this choice.  

Fig.~\ref{visc} (a) shows a characteristic feature of the colloidal glass transition: for $\phi=0.55$ and 0.58 below the colloidal glass transition, the viscosity is nearly uniform within a Newtonian regime at low $\Sigma/p$, while for $\phi\ge 0.6$ above the colloidal glass transition, the viscosity increases with decreasing $\Sigma/p$ until it becomes too large to measure in our simulations.  Outside of the Newtonian regime, the system exhibits shear thinning, with $\eta\sqrt{\sigma/pm}$ decreasing with increasing $\Sigma/p$.  The apparent shear thickening for the thermostatted simulations at very large $\Sigma/p$ is an artifact of the nonphysical layering transition.

Even restricting the calculations to the state points for which we conducted simulations longer than $500\tau$, the data for $\eta_{\rm EH}$ are much less precise.  However, two observations are clear.  First, the value of $\eta_{\rm EH}$ at given values of $\phi$ and $\Sigma/p$ is in most cases somewhat smaller for the thermostatted simulations than for the inelastic simulations.  This is the only quantity that we measure that is sensitive to the energy dissipation mechanism at low or moderate $\Sigma/p$.  We find that the discrepancy is accounted for by a somewhat faster relaxation of the shear stress correlation function in the thermostatted simulations compared to the inelastic simulations, though the time scales for the correlation function remain on the order of $\tau$ for both inelastic and thermostatted simulations.  Apparently, the profile-biased thermostat damps out some of the shear stress correlations that persist longer in the inelastic simulations.  Second, analogous to our result for the compressibility, the behavior of $\eta_{\rm EH}$ is markedly different than the behavior of $\eta$ at large $\Sigma/p$.  While $\eta\sqrt{\sigma/pm}$ uniformly decreases with increasing $\Sigma/p$, $\eta_{\rm EH}\sqrt{\sigma/pm}$ is non-monotonic for large $\phi$, first decreasing and then increasing with increasing $\Sigma/p$, and is roughly uniform for smaller $\phi$.

We define the shear viscosity temperature $T_\eta$ as the parameter that replaces $T$ to satisfy the Einstein-Helfand expression:
\begin{equation}
T_\eta\equiv \dfrac{1}{\eta}\lim_{t\rightarrow\infty}\dfrac{dH(t)}{dt}=\dfrac{\eta T_{\rm kin}}{\eta_{\rm EH}}.
\label{teta}
\end{equation}
Fig.~\ref{visc} (c) shows that $T_\eta$ behaves qualitatively similarly to $T_\chi$: for $\phi=0.55$ and 0.58, $T_\eta\simeq T_{\rm kin}$ at small $\Sigma/p$ and increases at large $\Sigma/p$, while for $\phi\ge 0.6$, $T_\eta$ is significantly larger than $T_{\rm kin}$ for all accessible values of $\Sigma/p$ and increases uniformly with $\Sigma/p$.

\begin{figure}
\begin{center}
\includegraphics[width=3in]{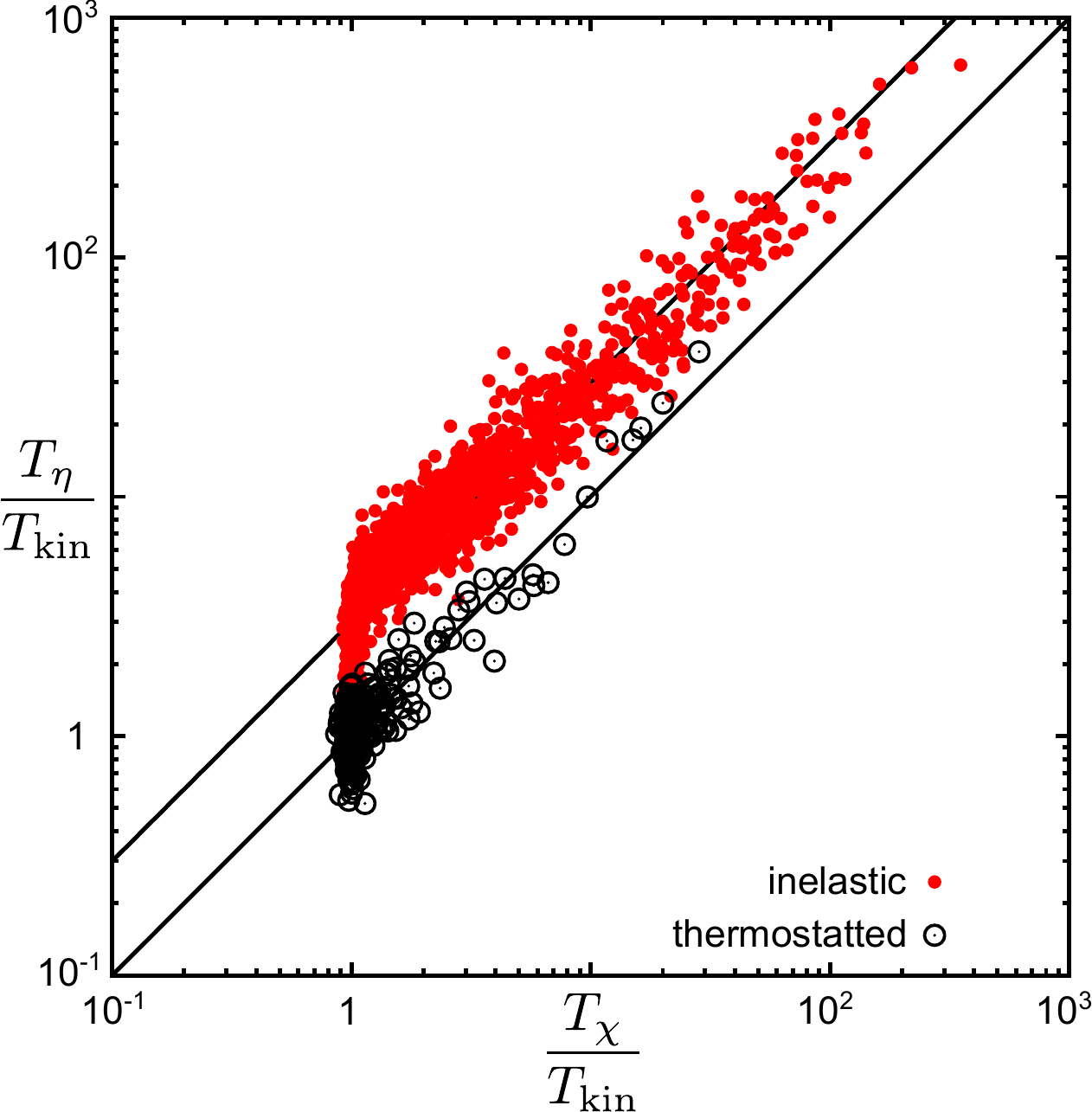}
\end{center}
\caption{Viscosity effective temperature vs compressibility effective temperature for all simulations that were run long enough to measure both quantities.  Red, open symbols are for thermostatted simulations, while black, filled symbols are for inelastic simulations.  The solid lines representing $T_\eta=3T_\chi$ and $T_\eta=T_\chi$ are visual guides.
}
\label{compare}
\end{figure}

In Fig.~\ref{compare}, we show that $T_\chi$ and $T_\eta$ represent a consistent value of the effective temperature.  We plot $T_\eta/T_{\rm kin}$ vs $T_\chi/T_{\rm kin}$ for a broad range of inelastic and thermostatted simulations within the range $0.2<\phi<0.636$ and $2 \times 10^{-4} <\Sigma/p<0.5$.  The points near $T_\chi/T_{\rm kin}=T_\eta/T_{\rm kin}=1$ represent near-equilibrium conditions, where the system is characterized by a single temperature, $T_\chi=T_\eta=T_{\rm kin}=T$.  For the thermostatted simulations, the viscosity and compressibility temperatures away from equilibrium are strongly correlated, clustered around the line $T_\eta=T_\chi$ representing a single effective temperature.  We are unable to test this relationship for thermostatted simulations beyond $T_\chi/T_{\rm kin}=30$, due to the layering transition at very large $\Sigma/p$.  However, for the inelastic simulations, we are able to explore the range $1<T_\chi/T_{\rm kin}<300$ without encountering artifacts.  Throughout this range, we find a strong correlation between $T_\eta$ and $T_\chi$.  However, the data clusters around a line representing $T_\eta=3T_\chi$, not $T_\eta=T_\chi$.  While we have already credited the discrepancy between thermostatted and inelastic simulations to the faster decay of shear stress correlations in the thermostatted simulations, we cannot explain the factor of 3 difference between $T_\eta$ and $T_\chi$.  However, given that the range (300) is much larger than the discrepancy (3), we find that $T_\eta$ and $T_\chi$ represent a consistent value of the effective temperature to a reasonable level of approximation.

\section{Effective temperature controls mobility}

\begin{figure}
\begin{center}
\includegraphics[width=3in]{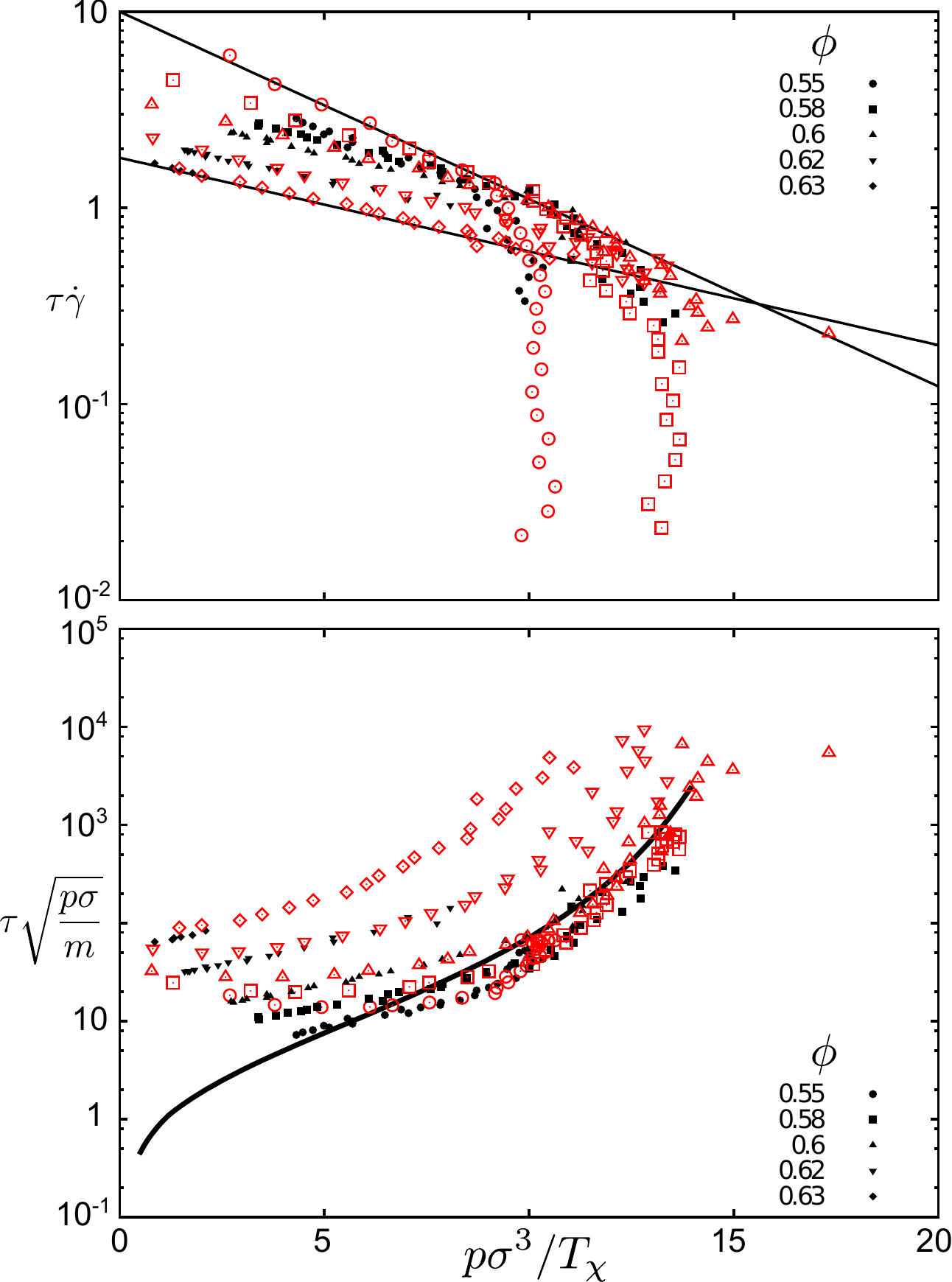}
\end{center}
\caption{(a) Yield strain $\tau\dot\gamma$ vs $p\sigma^3/T_\chi$.  The lines are visual guides satisfying the equations $\tau\dot\gamma=10\exp(-0.22p\sigma^3/T_\chi)$ and $\tau\dot\gamma=1.8\exp(-0.11p\sigma^3/T_\chi)$.  (b) Relaxation time nondimensionalized by the pressure $\tau\sqrt{p\sigma/m}$ vs $p\sigma^3/T_\chi$.  The solid curve is unsheared, equilibrium data for which $T_\chi=T$.  In each panel, data are presented for a selection of four packing fractions.  Red, open symbols are for thermostatted simulations, while black, filled symbols are for inelastic simulations.
}
\label{taurate}
\end{figure}

Having established that low-frequency density and shear stress fluctuations of sheared hard spheres are described by a consistent value of $T_{\rm eff}$, we now turn to establishing the role of $T_{\rm eff}$ in controlling the mobility, which we characterize by the relaxation time $\tau$.  In the following, we focus on $T_\chi$ as the more precise measure of the effective temperature.  Fig.~\ref{taurate} shows two different representations of the relationship between $\tau$ and $T_\chi$ for the same five packing fractions and range of shear stresses examined in Figs.~\ref{compress} and~\ref{visc}.  In order to show the relationship in a way that is independent of any arbitrary choice of units, we present the relationship in dimensionless form.  Because there are no internal energy scales in the hard-sphere interaction, the effective temperature can only be compared the external energy scales like $T_{\rm kin}$ and $p\sigma^3$.  Since we are interested both in cases where $T_\chi \simeq T_{\rm kin}$ and where $T_{\rm kin}$ is negligible, we compare $T_\chi$ to $p\sigma^3$ rather than $T_{\rm kin}$.
In Fig.~\ref{taurate} (a) and (b), we represent $\tau$ using two different dimensionless groups.  In Fig.~\ref{taurate} (a), we plot the the average yield strain $\tau\dot\gamma$, the average amount of strain built up locally between relaxations.  In Fig.~\ref{taurate} (b), we plot $\tau\sqrt{p\sigma/m}$, the relaxation time nondimensionalized by the pressure.  The time scale $\sqrt{m/p\sigma}$ is the pressure-driven rearrangement time that sets the time scale for a loose configuration to be compressed into a close-packed configuration due to the pressure.  Scaling both $\tau$ and $T_{\rm eff}$ by the pressure in Fig.~\ref{taurate} (b) serves to capture not only the behavior of the hard sphere system, but also the behavior of the related soft sphere system to leading order in $p\sigma^3/\epsilon$, where $\epsilon$ is the energy scale of the soft repulsion~\cite{Xu2009b, Haxton2011, Schmiedeberg2011}.

\begin{figure*}
\begin{center}
\includegraphics[width=6in]{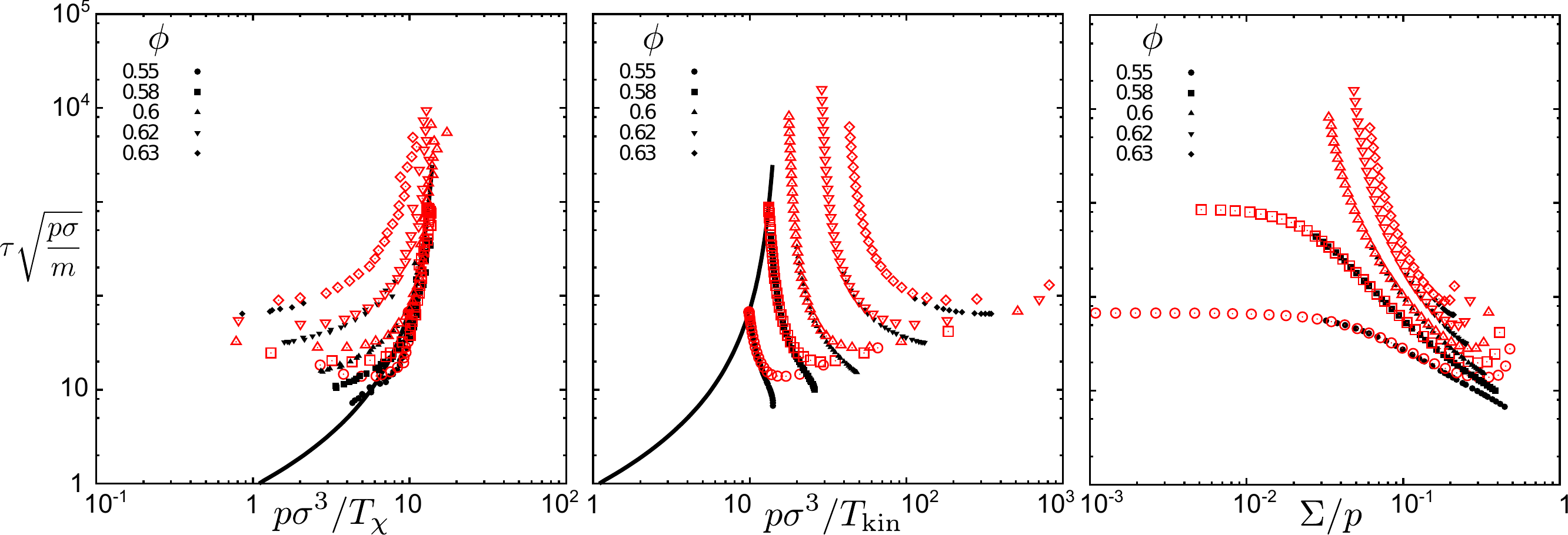}
\end{center}
\caption{Relaxation time $\tau\sqrt{p\sigma/m}$ vs (a) inverse compressibility temperature $p\sigma^3/T_\chi$, (b) inverse kinetic temperature $p\sigma^3/T_{\rm kin}$, and (c) shear stress $\Sigma/p$.  Data are presented for a selection of four packing fractions.  Red, open symbols are for thermostatted simulations, while black, filled symbols are for inelastic simulations.  The solid curve that appears in panels (a) and (b) is unsheared, equilibrium data for which $T_\chi=T$.  In each panel, data are presented for a selection of four packing fractions.}
\label{tau}
\end{figure*}

In Fig.~\ref{taurate} (a) we examine the relationship between the effective temperature and the mechanical state of the system by plotting the average yield strain vs $p\sigma^3/T_\chi$ for our selection of five packing fractions.  We find a crossover between two regimes.  In what we call the \textit{solid response} regime, at low $p\sigma^3/T_\chi$ and high $\tau\dot\gamma$, the data at each packing fraction follows the relationship
\begin{equation}
\tau\dot\gamma\simeq c_0\exp(-c_1p\sigma^3/T_\chi),
\label{exp}
\end{equation}
which appears as a linear relationship on the log-linear scale of Fig.~\ref{taurate} (a).  As illustrated by the visual guides in Fig.~\ref{taurate} (a), the coefficient $c_0$ ranges from 1.8 for $\phi=0.63$ to 10 for $\phi=0.55$, and $c_1$ ranges from 0.11 for $\phi=0.63$ to 0.22 for $\phi=0.55$.  The appreciable average yield strain and its dependence on $T_\chi$ expressed by Eq.~\ref{exp} indicates that within the solid-response regime, the system behaves like a continuously deformed solid with a mechanical response controlled by the internal state parameter $T_\chi/p\sigma^3$.  At high $p\sigma^3/T_\chi$ and low $\tau\dot\gamma$, the data for each packing fraction peel off of the linear relationship described by Eq.~\ref{exp}.  We will show shortly that these data correspond to a fluid response.  The vertical parts of the data for $\phi=0.55$ and 0.58 indicate that for $\phi$ below the colloidal glass transition, the system approaches the equilibrium limit of $\dot\gamma\rightarrow 0$ and $p\sigma^3/T_\chi \rightarrow p\sigma^3/T$.  We note that the plot of $\Sigma/p$ vs $p\sigma^3/T_\chi$ is qualitatively similar to Fig.~\ref{taurate} (b), due to the fact that the shear modulus in a Maxwell-fluid model for hard spheres is roughly proportional to the pressure~\cite{Haxton2011}.  This is consistent with our result in Ref.~\cite{Haxton2007} that $\Sigma\propto\exp(-\Delta E/T_{\rm eff})$ for soft, repulsive disks above random close packing; the energy scale $\Delta E$ is controlled by the pressure.

The form of Eq.~\ref{exp} is consistent with the shear transformation zone theory that describes plasticity of amorphous solids~\cite{Falk2011}.  In the theory, the effective temperature describes the slow, configurational degrees of freedom associated with rearrangements of neighboring particles, and plastic rearrangements occur in zones whose internal configurational degrees of freedom are modeled as a system with two orientational states.  The steady-state solution to the theory at low temperature is
$\dot\gamma=(1/2)\epsilon_0N_{\rm Z}\rho_{\rm Z}R(\Sigma, T),$
where $\epsilon_0$ is the strain per transformation, of order 1, $N_{\rm Z}$ is the number of particles in a zone, $\rho_{\rm Z}=\exp(-E_{\rm Z}/T_{\rm eff})$ is the density of zones, $E_{\rm Z}$ is the energy to create a zone, and $R(\Sigma, T)$ is the rate of transformations in a zone~\cite{Falk2011}.  Without making any assumption about the form of $R(\Sigma, T)$, we assume that it is proportional to the diffusion coefficient; that is, we assume that the frequency of plastic rearrangements is linked to the frequency at which spheres make large displacements.  This leads to
$\dot\gamma\tau=(1/2)\tilde R\epsilon_0N_{\rm Z}\exp(-E_{\rm Z}/T_{\rm eff}),$
where $\tilde R$ is the average number of times a zone rearranges per relaxation time $\tau$.  Comparing to Eq.~\ref{exp} yields values of $\tilde R\epsilon_0N_{\rm Z}$ between 3.6 and 20 and values of $E_{\rm Z}$ between $0.11p\sigma^3$ and $0.21p\sigma^3$; that is, the free volume associated with creating a shear transformation zone is approximately $0.11\sigma^3$ to $0.22\sigma^3$.  Our measured excitation energies $E_{\rm Z}$ are similar to the characteristic energy scales extracted from the dependence of $\tau\sqrt{p\sigma/m}$ on $T/p\sigma^3$ for equilibrium hard spheres, for which fits of the Vogel-Fulcher and Elmatad-Chandler-Garrahan~\cite{Elmatad2009, Keys2011} forms yield energy scales of $0.18p\sigma^3$ and $0.25p\sigma^3$, respectively~\cite{Xu2009b}.

In Fig.~\ref{taurate} (b) we demonstrate that for parameters outside of the solid response regime, the relaxation time is controlled by the effective temperature in a characteristically fluid way.  In Fig.~\ref{taurate} (b) we plot $\tau$ vs $T_\chi$, where each quantity is nondimensionalized by the pressure, for the same set of data as in Fig.~\ref{taurate} (a).  Fig.~\ref{taurate} (b) is a standard Arrhenius plot adapted to the effective temperature, a log-linear plot of $\tau\sqrt{p\sigma/m}$ versus the ratio of pressure to effective temperature, $p\sigma^3/T_\chi$.  For each packing fraction, the shear stress decreases to the right and upward; relaxation time increases and effective temperature decreases as the shearing decreases.  For comparison, the solid curve represents equilibrium simulations under no shear, which possess a single temperature $T_\chi=T_{\rm kin}=T$.  In equilibrium, the relaxation time depends on the temperature according to the dimensionless function represented by the solid curve,
\begin{equation}
\tau\sqrt{p\sigma/m}=f(T/p\sigma^3).
\label{eqeq}
\end{equation}
The packing fraction $\phi$ increases as this equilibrium curve proceeds to the right and upward; the relaxation time increases as $\phi$ approaches the colloidal glass transition and $p\sigma^3/T$ increases according to the equation of state.  We find that outside of the solid response regime--wherever the data in Fig.~\ref{taurate} (a) does not follow Eq.~\ref{exp}--the data clusters around the equilibrium curve.  Within this \textit{fluid response} regime, the dependence of relaxation time on the two parameters $\phi$ and $\Sigma/p$ is captured by the equilibrium expression, Eq.~\ref{eqeq}, with $T$ replaced by $T_\chi$.  This suggests that within the fluid response regime, the same mechanism that controls the relaxation of unsheared hard spheres also controls the relaxation of sheared hard spheres: spheres only flow if low-frequency fluctuations and large enough compared to the pressure to open up a sufficient amount of free volume~\cite{Xu2009b}.  However, for the sheared hard spheres, the low-frequency fluctuations are characterized by the effective temperature rather than the kinetic temperature.


In order to more clearly establish that the effective temperature, not the kinetic temperature, controls the mobility in the fluid response regime, we compare the dependence of $\tau\sqrt{p\sigma/m}$ on $p\sigma^3/T_\chi$ with its dependence on $p\sigma^3/T_{\rm kin}$ in Fig.~\ref{tau} (a) and (b), respectively.  To aid the comparison, we plot $p\sigma^3/T_\chi$ and $p\sigma^3/T_{\rm kin}$ on the same logarithmic scale.  Note that Fig.~\ref{tau} (a) is a log-log version of the log-linear Fig.~\ref{taurate} (b).  For reference, Fig.~\ref{tau} (c) shows the same data plotted vs $\Sigma/p$.   The equilibrium curve that appears in Fig.~\ref{tau} (a) and (b) is identical, due to the fact that $T_\chi=T_{\rm kin}$ in equilibrium.  While much of the data--including large portions of the $\phi=0.55$, 0.58, and 0.6 data and some of the $\phi=0.62$ data--cluster near the equilibrium curve in Fig.~\ref{tau} (a), the sheared data in Fig.~\ref{tau} (b) only approach the equilibrium curve in the limit $\Sigma/p\rightarrow 0$ for $\phi=0.55$ and 0.58.  For the packing fractions above the colloidal glass transition, where the near-equilibrium condition $\Sigma/p<<1$ is never met (see Fig.~\ref{tau} (c)), the sheared data remain distinct from the equilibrium curve.

\section{Effective temperature and the jamming phase diagram}

\begin{figure}
\begin{center}
\includegraphics[width=3in]{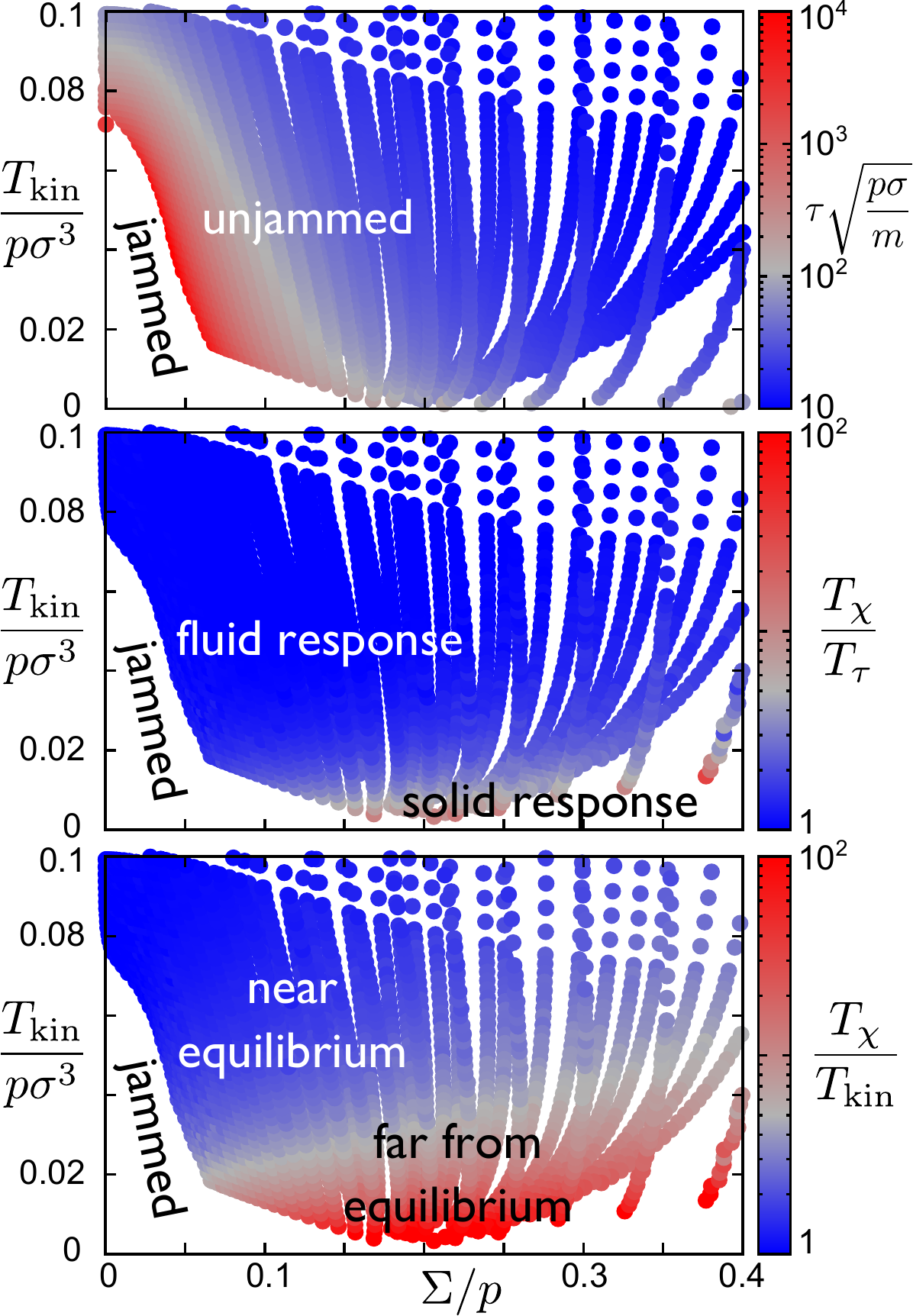}
\end{center}
\caption{Color plots of (a) $T_\chi/T_{\rm kin}$ and (b) $\tau\sqrt{p\sigma/m}$ as functions of kinetic temperature $T_{\rm kin}/p\sigma^3$ and shear stress $\Sigma/p$.
}
\label{splot}
\end{figure}

We have shown that the ratio of effective temperature to pressure, $T_\chi/p\sigma^3$, controls the mobility of sheared hard spheres according to two different mechanisms in two different regimes.  In the solid response regime, the effective temperature controls the mechanical state of the system, as evidenced by the dependence of the average yield strain.  In the fluid response regime, the effective temperature facilitates relaxation by doing work against the pressure, analogous to how the temperature facilitates relaxation in the unsheared system.  It is instructive to organize these regimes within the jamming phase diagram~\cite{Liu1998, Haxton2011} that describes the relaxation of repulsive spheres as a function of temperature, packing fraction, and applied stress.  Ref.~\cite{Haxton2011} recast the jamming phase diagram in dimensionless form in terms of the parameters $T_{\rm kin}/p\sigma^3$, $\Sigma/p$, and $p\sigma^3/\epsilon$, where $\epsilon$ is the energy scale characterizing the repulsive interaction.  For hard spheres, $\epsilon=\infty$ and the jamming phase diagram reduces to the hard-sphere limit, $p\sigma^3/\epsilon\rightarrow 0$.  In this limit the relaxation depends on two parameters, an equilibrium parameter $T_{\rm kin}/p\sigma^3$ characterizing the strength of thermal fluctuations and a nonequilibrium parameter $\Sigma/p$ characterizing the strength of shearing.

Fig.~\ref{splot} locates the solid response, fluid response, near equilibrium, and jammed regimes in the dimensionless jamming phase diagram.  The three panels in Fig.~\ref{splot} represent the same sets of simulations, a large collection of all inelastic and thermostatted simulations for which $\Sigma/p<0.4$ and $T_{\rm kin}/p\sigma^3<0.1$.  Each panel presents a color map of a different quantity.  Fig.~\ref{splot} (a) is a color map of the dimensionless relaxation time $\tau\sqrt{p\sigma/m}$.  The region where data appears is the unjammed region of the phase diagram, where the equilibration and relaxation times are short enough that we can run simulations to calculate $\tau$.  The red region indicates very large relaxation times that are near the limits of our simulations.  The dynamic jamming transition, defined as the locus of points for which $\tau\sqrt{p\sigma/m}$ equals some large but arbitrary number (like $10^4$), separates the jammed region at small $\Sigma/p$ and $T_{\rm kin}/p\sigma^3$ from the unjammed region at large $\Sigma/p$ and/or large $T_{\rm kin}/p\sigma^3$.

Fig.~\ref{splot} (b) shows that the unjammed region includes both the fluid response and the solid response regimes.  Fig.~\ref{splot} (b) is a color plot of the ratio of the effective temperature $T_\chi$ to the the temperature $T_\tau\equiv p\sigma^3f^{-1}(\tau\sqrt{p\sigma/m})$ defined by inverting the equilibrium (zero shear) relationship, Eq.~\ref{eqeq}, between relaxation time and temperature.  A value of $T_\chi/T_\tau=1$ indicates that the relaxation time is predicted by Eq.~\ref{eqeq} with $T$ replaced by $T_\chi$, corresponding to a collapse onto the equilibrium curve in Fig.~\ref{taurate} (b) and Fig.~\ref{tau} (a).  Notice that a large portion of the jamming phase diagram is blue, demarcating the fluid response regime where $T_\chi/T_\tau$ is equal to or not much larger than 1 and the relaxation time is reasonably well predicted by Eq.~\ref{eqeq} with $T$ replaced by $T_\chi$.  In particular, much of the dynamic jamming transition occurs within the fluid response regime.

A comparison with Fig.~\ref{splot} (c) emphasizes that fluid response does \textit{not} imply near-equilibrium conditions.  Fig.~\ref{splot} is a color plot of the ratio of the effective temperature to the kinetic temperature.  Values near 1, represented by blue, indicate that the system is near equilibrium, with fast and slow degrees of freedom characterized by a single temperature.  Values much greater than 1 indicate that the system is strongly out of equilibrium, with an effective temperature much greater than the temperature of the high frequency velocity fluctuations.  While a large portion of the unjammed region is within the near-equilibrium regime, this regime is distinctly smaller than the fluid response regime of Fig.~\ref{splot} (b).

Taken together, the three panels of Fig.~\ref{splot} show that while the dynamic jamming transition at nonzero shear stress is largely controlled by a competition between low frequency fluctuations and pressure in the same way as at zero shear stress, these low-frequency fluctuations are characterized by an effective temperature that may be much larger than kinetic temperature.

The jamming phase diagram provides a framework for applying our hard-sphere results to systems with soft interactions.  Moving away from the hard-sphere limit, the relaxation of soft, repulsive spheres varies continuously with the parameter $p\sigma^3/\epsilon$ that characterizes the softness of the potential~\cite{Xu2009b, Haxton2011}.  Even far from the hard-sphere limit, where the relaxation time is not adequately approximated by its form at $p\sigma^3/\epsilon=0,$ the dependence of $\tau\sqrt{p\sigma/m}$ on $T/p\sigma^3$ (with no shear) can be mapped onto the hard-sphere behavior by computing an effective hard-sphere diameter from the structure of the soft-sphere system~\cite{Schmiedeberg2011}.  This suggests that while the numeric form of the relaxation changes far away from $p\sigma^3/\epsilon=0$, the mechanisms remain the same.  We therefore expect that for $p\sigma^3/\epsilon>0$ the effective temperature continues to facilitate relaxation via the same two mechanisms as in the hard sphere limit.  While the boundaries between the two mechanisms may shift at elevated  $p\sigma^3/\epsilon$, we expect that the solid response will continue to dominate at low $T_{\rm kin}/p\sigma^3$ and high $\Sigma/p$ while the fluid response dominates elsewhere.  Analysis of our earlier simulations~\cite{Haxton2007} of sheared two-dimensional soft disks in terms of dimensionless quantities is consistent with this picture.


\section{Conclusions}

We have shown that the ratio of effective temperature to pressure, $T_{\rm eff}/p\sigma^3$, controls the mobility of sheared hard spheres throughout the unjammed region of the jamming phase diagram spanned by packing fraction and applied shear stress.  The effective temperature characterizes both the amplitude of static density fluctuations relative to the compressibility and the amplitude of low-frequency shear stress fluctuations relative to the shear viscosity.  At high shear stress and low kinetic temperature, the effective temperature departs significantly from the kinetic temperature that characterizes high-frequency fluctuations.  In this far-from-equilibrium regime, the effective temperature, not the kinetic temperature, controls the diffusive motion of the spheres.  

We find that the mechanism by which $T_{\rm eff}/p\sigma^3$ controls the mobility depends on where the system lies in the jamming phase diagram, crossing over between two regimes.  For very low values of the kinetic temperature relative to the pressure, the sheared hard spheres respond like a solid.  They maintain an average local yield strain that depends on the value of $T_{\rm eff}/p\sigma^3$.  The functional form of this dependence is consistent with the effective temperature's thermodynamic role in the solid-based shear transformation zone theory, in which the effective temperature controls the density of localized zones susceptible to plastic deformation.  In contrast, for moderate or large values of the kinetic temperature, the sheared hard spheres respond like a fluid.  The relaxation time depends on temperature in the same way as does the unsheared, equilibrium system, except that the relevant temperature is the effective temperature, \textit{not} the kinetic temperature.  This suggests that, just as the temperature facilitates flow in the unsheared system by doing work against the pressure, the effective temperature mobilizes the sheared spheres by working against the pressure to open up free volume.

By expressing our results in relation to the dimensionless version of the jamming phase diagram, we have illustrated how these mechanisms likely apply for systems with soft repulsive potentials.  It remains to be seen to what extent these mechanisms hold for systems with attractions.

\section{Acknowledgments}

I thank Andrea Liu for instructive discussions.  This work was funded by DOE DE-FG02-05ER46199 and DOE DE-AC02-05CH11231.

\end{document}